\documentclass{appolb}
\usepackage{graphicx}
\begin{document}
\eqsec  % uncomment this line to get equations numbered by (sec.num)
\title{Sensitivity of jet observables to the presence of quasi-particles in QGP%
\thanks{Presented by Z. Hulcher at Quark Matter 2022, Krakow, Poland}%
}
\author{Zachary Hulcher$^1$, Daniel Pablos$^2$, Krishna Rajagopal$^3$
\address{$^1$Stanford University, 450 Serra Mall, Stanford, CA, USA 94305}
\\[-3mm]
%Daniel Pablos %
\address{$^2$INFN, Sezione di Torino, via Pietro Giuria 1, I-10125 Torino, Italy}
\\[-3mm]
%Krishna Rajagopal
\address{$^3$MIT CTP, 77 Massachusetts Ave, Cambridge, MA, USA 02139}
}
\maketitle
\begin{abstract}
QGP, a strongly coupled liquid when viewed at length scales of ${\cal O}(1/T)$, must reveal quark- and gluon-like quasi-particles when probed with sufficiently high momentum transfer, since QCD is asymptotically free. High energy jet partons traversing the droplet of QGP produced in a heavy ion collision can trigger these high-momentum exchanges with medium constituents and so have the potential to reveal the presence of such quasi-particles, a key step toward the experimental study of the microscopic structure of QGP. 
We implement this physics within the hybrid strong/weak coupling model which, prior to this work, only accounted for nonperturbative aspects of parton energy loss. Elastic Moli\`ere scatterings between partons from a jet shower and medium  quasi-particles result in deflection of the propagating jet partons and struck thermal medium partons recoiling at large angles. 
We discuss the effect of Moli\`ere scatterings on some of the most widely used groomed and ungroomed jet substructure observables. Given the large impact on jet observables of the wakes generated by the hydrodynamic response of the QGP fluid as the shower and scattered partons lose energy and momentum to it, as well as the presence of selection biases, finding distinctive signatures of Moli\`ere scattering, and hence the presence of quasi-particles in QGP, is a challenge. Toward this end, we emphasize the discovery potential of subjet (jets within jets) distributions.

\end{abstract}

\section{The Hybrid Model with Moli\`ere Scatterings}

The hybrid strong/weak coupling model~\cite{Casalderrey-Solana:2014bpa,Casalderrey-Solana:2015vaa} combines the perturbative, high-$Q^2$, DGLAP evolution of jet dynamics with the nonperturbative interaction between partons in a jet shower and strongly coupled QGP. Indeed, the most probable interactions between a jet parton and the expanding cooling droplet of QGP produced in a heavy ion collision involve low-momentum transfers of order the QGP temperature $T$, which is comparable to the QCD confinement scale, and are thus nonperturbative. To describe this, we use holographic results for the energy loss of an energetic parton within strongly coupled plasma with a holographic description, derived at infinite coupling and large $N_c$~\cite{Chesler:2014jva,Chesler:2015nqz}. The energy loss rate $dE/dx$ for an energetic parton that has traveled a distance $x$
takes the form 
$(E_{\rm in}/x_{\rm stop})f(x/x_{\rm stop})$ where $f$ is known~\cite{Chesler:2014jva,Chesler:2015nqz} and
where $x_{\rm stop}\equiv (E_{\rm in}^{1/3}/{T}^{4/3})/2 \kappa_{\rm sc}$ is the maximum distance the parton can travel within the plasma before completely thermalizing. $\kappa_{\rm sc}$ is an $\mathcal{O}(1)$ parameter fixed by fitting to hadron and jet suppression measured at the LHC~\cite{Casalderrey-Solana:2018wrw}. The energy and momentum lost by the jet partons hydrodynamizes and 
becomes a wake in the expanding cooling droplet of QGP, also described hydrodynamically, 
that later decays into soft hadrons at the freeze-out hypersurface. We estimate the distributions of those hadrons by applying the Cooper-Frye prescription to the jet-induced perturbations, assuming that the background fluid is described by Bjorken flow and that the perturbations stay localized in rapidity around the jet~\cite{Casalderrey-Solana:2016jvj}. %

Our starting point for adding perturbative Moli\`ere scattering into the model
is the calculation~\cite{DEramo:2018eoy} of Moli\`ere scattering probabilities for an incident hard parton on a
``brick'' of length $L$ of quarks and gluons in thermal equilibrium at temperature $T$, see Fig.~1. Each relevant Feynman diagram yields a contribution of the form~\cite{DEramo:2018eoy}
	\begin{equation}
	\langle(n)\rangle_{DB}=\frac{1}{2(4\pi)^3}\frac{p\sin\theta}{p_{\rm in}|\mathbf{p}-\mathbf{p_{\rm in}}|T }\int dk_Tn_D(k_T)(1\pm n_B(k_{\chi}))\int d\phi \frac{1}{g_s^4}|M^{(n)}(t,u)|^2\,.
	\label{nDB}
	\end{equation}
Here,
$p_{\rm in}$ is the momentum of the incoming parton in the rest frame of the brick, $p$ and $\theta$ are the momentum and the angle (with respect to $\vec{p}_{\rm in}$) of the outgoing parton that is detected, $k_T$ is the momentum of the thermal parton $D$ that was struck, $k_{\chi}$ is the momentum of the outgoing parton $B$ that is not detected, 
$n_D(k)$ and $n_B(k)$ are Fermi-Dirac or Bose-Einstein distributions, as appropriate, the sign in front of $n_B$ is chosen accordingly,
$g_s$ is the strong coupling constant (we set $g_s=2.25$ throughout), $\phi$ is the azimuthal angle of the thermal parton around $p_{\rm in}$, 
$t$ and $u$ are Mandelstam variables, and $|M^{(n)}|^2$ is the massless 2-to-2 amplitude-squared for the process described by Feynman diagram $(n)$. See Ref.~\cite{DEramo:2018eoy} for expressions for the $M^{(n)}$ and for details on how to count and add up all the terms like Eq.~(\ref{nDB}) to construct scattering probabilities, for example the probability density $F^{G\rightarrow {\rm all}}(p,\theta)$ for an incident gluon to undergo a Moli\`ere scattering that yields a gluon, quark or antiquark with momentum $p$ and angle $\theta$ plotted in Fig.~1.

\begin{figure}[t!]
\centering
\includegraphics[width=.49\textwidth]{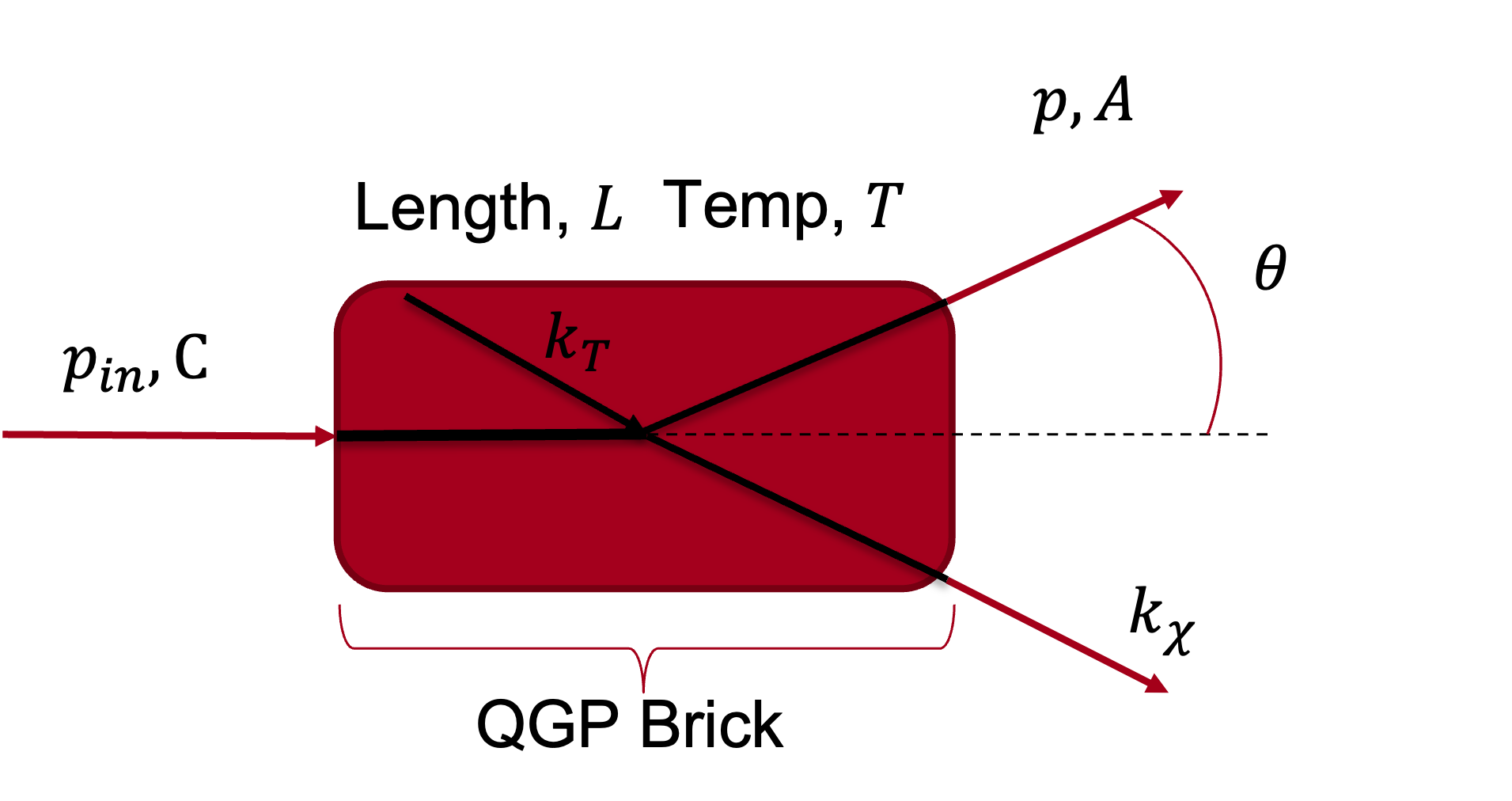}
\includegraphics[width=.49\textwidth]{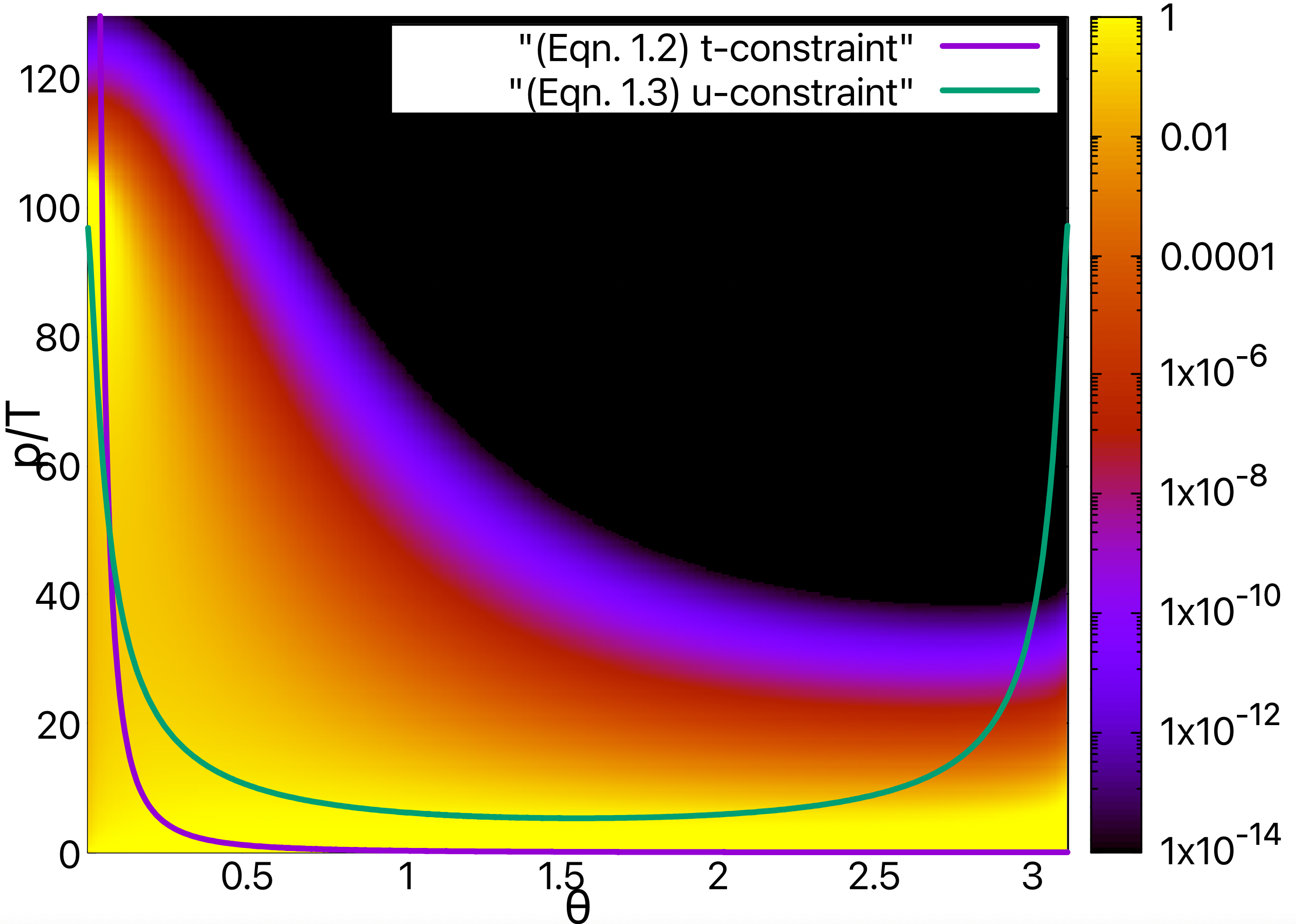}
\caption{Left: cartoon of Moli\`ere scattering of a parton $C$ with momentum $p_{\rm in}$ incident on a brick of length $L$  and temperature $T$ scattering off a thermal parton $D$ with momentum $k_T$ and yielding an observed parton $A$ with momentum $p$ and an unobserved parton $B$ with momentum $k_\chi$~\cite{DEramo:2018eoy}. Right: Preliminary result for $F^{G\rightarrow {\rm all}}(p,\theta)$ (depicted via color) for 
$p_{\rm in}=100 T$ and $L=15/T$. The $t$-constraint from Eq.~(\ref{eq:t_and_u_constraints}) is shown, as is the approximation (\ref{eq:uconstraintapprox}) to the $u$-constraint.  A full treatment of the $u$-constraint cannot be represented as a curve on this plot.}
\label{Fig:Fig1}
\end{figure}

It would be a mistake to extend our calculation of 2-to-2  Moli\`ere scattering
off an individual quasi-particle in the QGP down to low momentum transfer, where multiple soft scatterings dominate in weakly coupled treatments or 
where, in a strongly coupled approach, one finds (quantitatively small) Gaussian transverse momentum broadening~\cite{Casalderrey-Solana:2016jvj} 
as well as the energy loss that we are already taking into account via the strongly coupled energy loss rate.
To avoid double-counting (and mistreating) this regime, we shall include only
those scatterings
in which the momentum transfer is larger than 
$2 m_D \simeq 5.5 T$. Specifically, we require that
	\begin{equation}
|t|=2p_{\rm in}p(1-\cos\theta)>a^2 m_D^2 \, , \quad
	|u|=2(p_{\rm in}k_{\chi}-\vec{p}_{\rm in}\cdot \vec{k}_{\chi})>a^2 m_D^2,
	\label{eq:t_and_u_constraints}
\end{equation}
where $a$ is a number we take to be 2. 
The $t$-constraint keeps both $\theta$ and $p$ from becoming too small, see Fig.~1.
Noting that the thermal momentum $k_T$ is of order $T$, if we assume that $k_T\ll p_{\rm in}$ and $\ll p$, then the $u$-constraint can be recast as a constraint in the $(p,\theta)$ plane that takes the form~\cite{DEramo:2018eoy}
\begin{equation}
    |u|\simeq 2 p_{\rm in} 
    \left[ \left( p_{\rm in} - p \right) - \sqrt{ \left( p_{\rm in} - p \right)^2  - \left( p \sin \theta \right)^2} \ \right] > a^2 m_D^2\ ,
    \label{eq:uconstraintapprox}
\end{equation}
which serves to require that $p > a m_D$ at all $\theta$ and that keeps $\theta$ from getting too close to either 0 or $\pi$, see Fig.~1.
If we were to choose $a\gg 1$, the expression (\ref{eq:uconstraintapprox}) would indeed be a good approximation (since thermal particles with 
$k_T$ of order $p$ would then be rare).
However, we are choosing $a=2$. This means that although we shall always choose $p_{\rm in}\gg T\sim k_T$ we must consider the possibility that $p\sim k_T$. If $p=k_T$, then by energy 
conservation we also have $p_{\rm in}=k_\chi$ and by momentum conservation we then have $\vec{p}=\vec{k}_T$ and $\vec{p}_{\rm in}=\vec{k}_\chi$, which means zero momentum transfer, $u=0$, see (\ref{eq:t_and_u_constraints}). 
This means that when performing the $k_T$ integration in (\ref{nDB}) we must exclude
a region in the space of kinematic variables $(p,\theta,k_T,\phi)$ around $k_T=p$ where $|u|< a^2 m_D^2$.
This means that the full implications of the $u$-constraint cannot be depicted as a curve on the $(p,\theta)$ plane of Fig.~1.  A calculation in which we implement the $u$-constraint precisely is in progress.  
For the preliminary results shown in Fig.~1, we have implemented the $u$-constraint somewhat crudely via separate cuts in $p$, $\theta$ and  $k_\perp$.

We then employ the calculation of the probability of Moli\`ere scattering on a brick of QGP  as follows. We treat every parton in the developing hybrid model jet shower at every time-step $d\tau$ as incident on a brick of plasma
of thickness $d\tau$ with a temperature $T$ corresponding to the local temperature at that spacetime point. Most often, no Moli\`ere scattering occurs.
Rarely, with a probability that we have calculated, a medium parton with momentum $\vec k_T$ scatters off the jet parton yielding two partons with momenta $\vec p$ and $\vec k_\chi$.
When this happens, we immediately check whether the 
$u$-constraint (\ref{eq:t_and_u_constraints}) is actually satisfied. Because of the imprecision of our current implementation, sometimes $u$ is too small and if so we reject the scattering.
Implementing the $u$-constraint with imprecise cuts followed by this sampling procedure means that we sample slightly less phase space than we should, meaning that we are slightly underestimating the contribution of Moli\`ere scattering.
After a Moli\`ere scattering, we treat both outgoing partons as components of the jet shower henceforth. Both lose energy nonperturbatively from then on and, although unlikely, either could scatter again.

\section{Impact of Moli\`ere Scatterings on Jet Observables}

\begin{figure}[t!]
\centering
\includegraphics[width=0.49\textwidth]{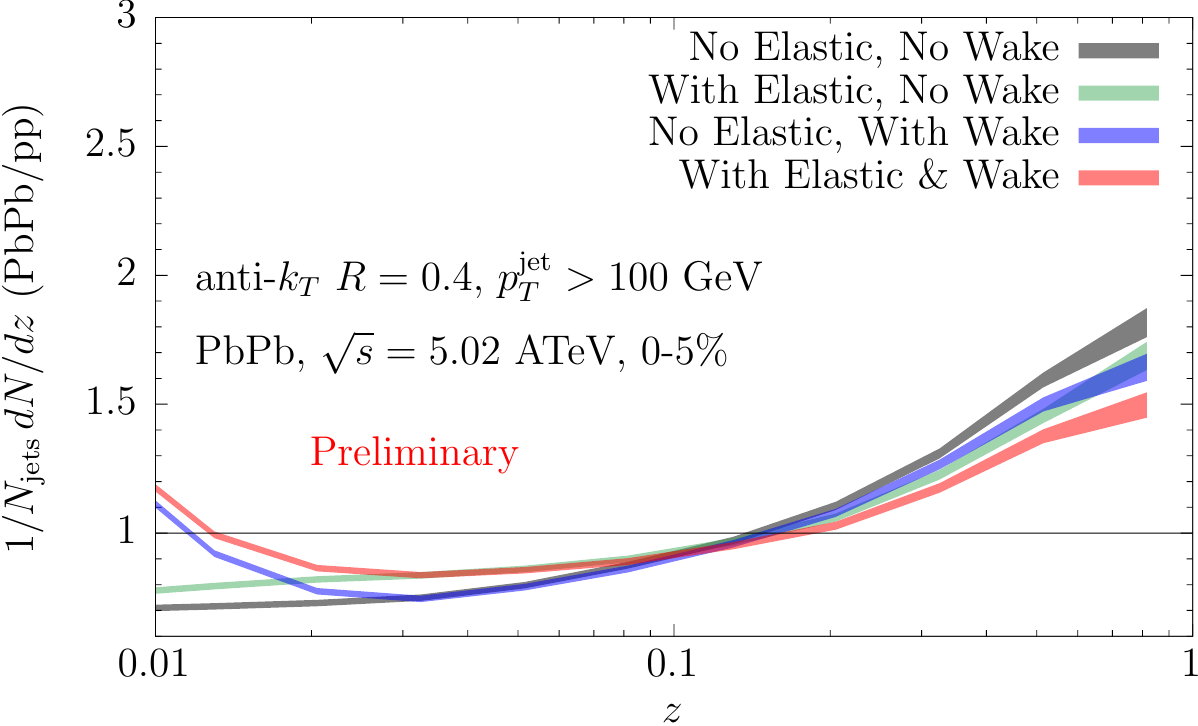}
\includegraphics[width=0.49\textwidth]{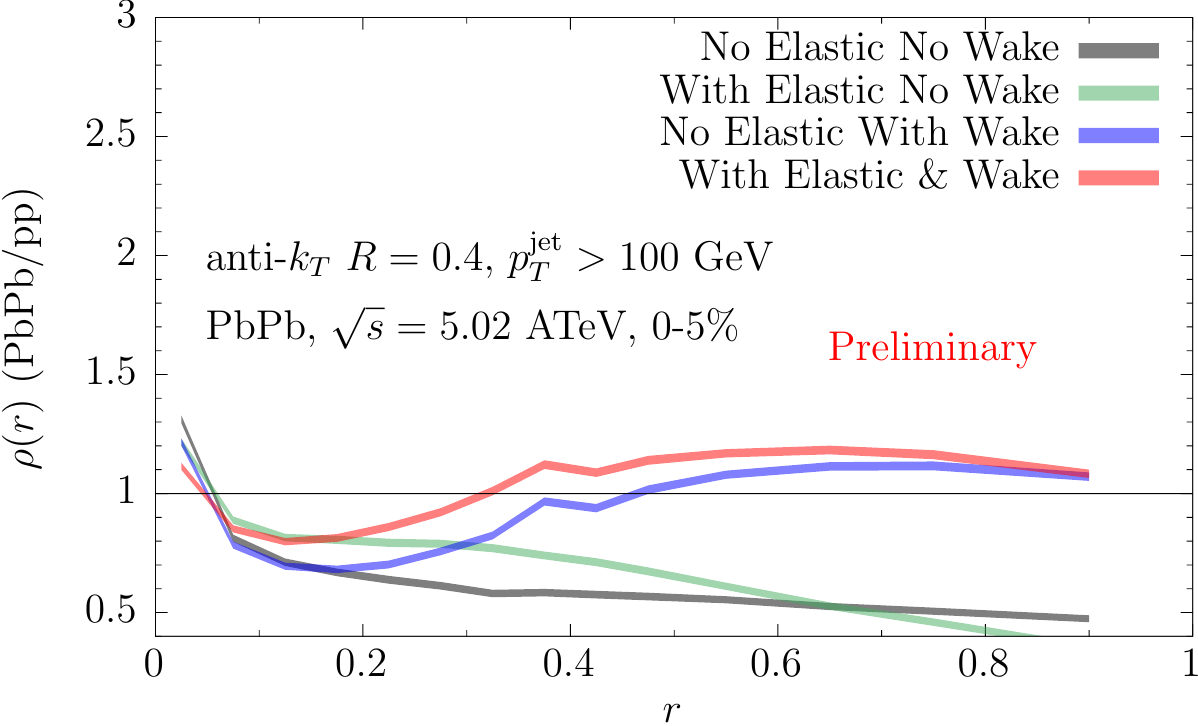}

\vspace{0.07in}

\includegraphics[width=0.49\textwidth]{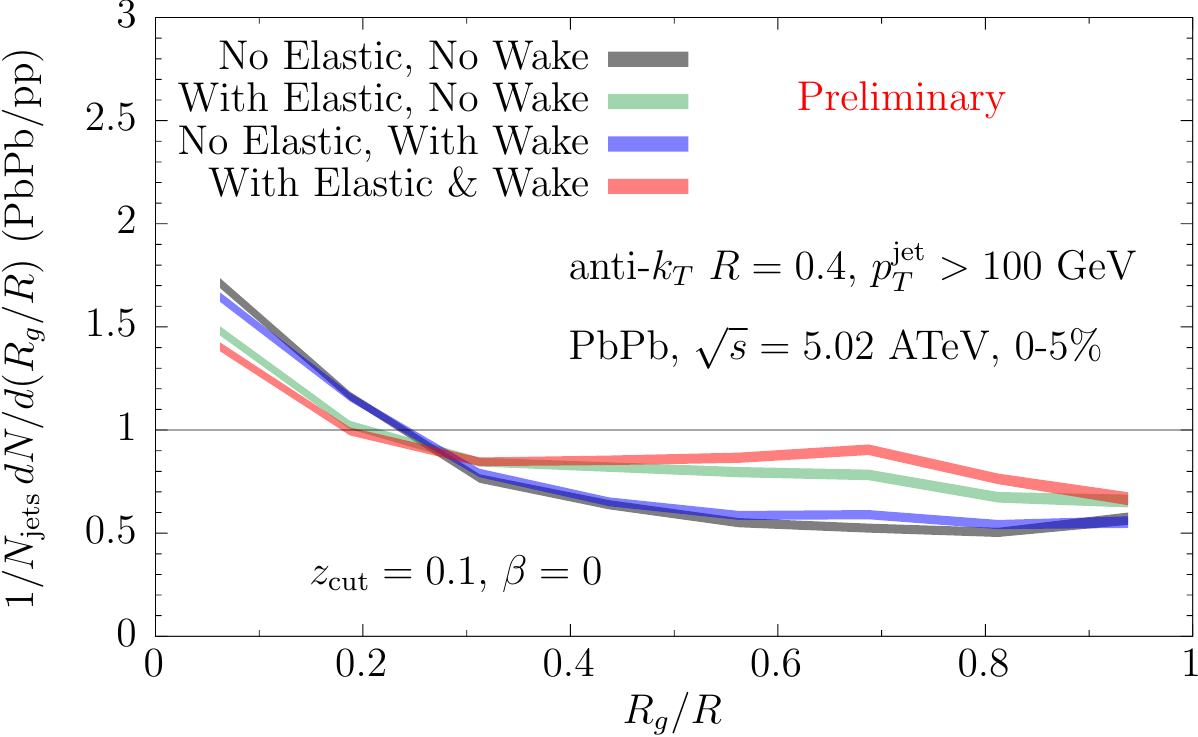}
\includegraphics[width=0.49\textwidth]{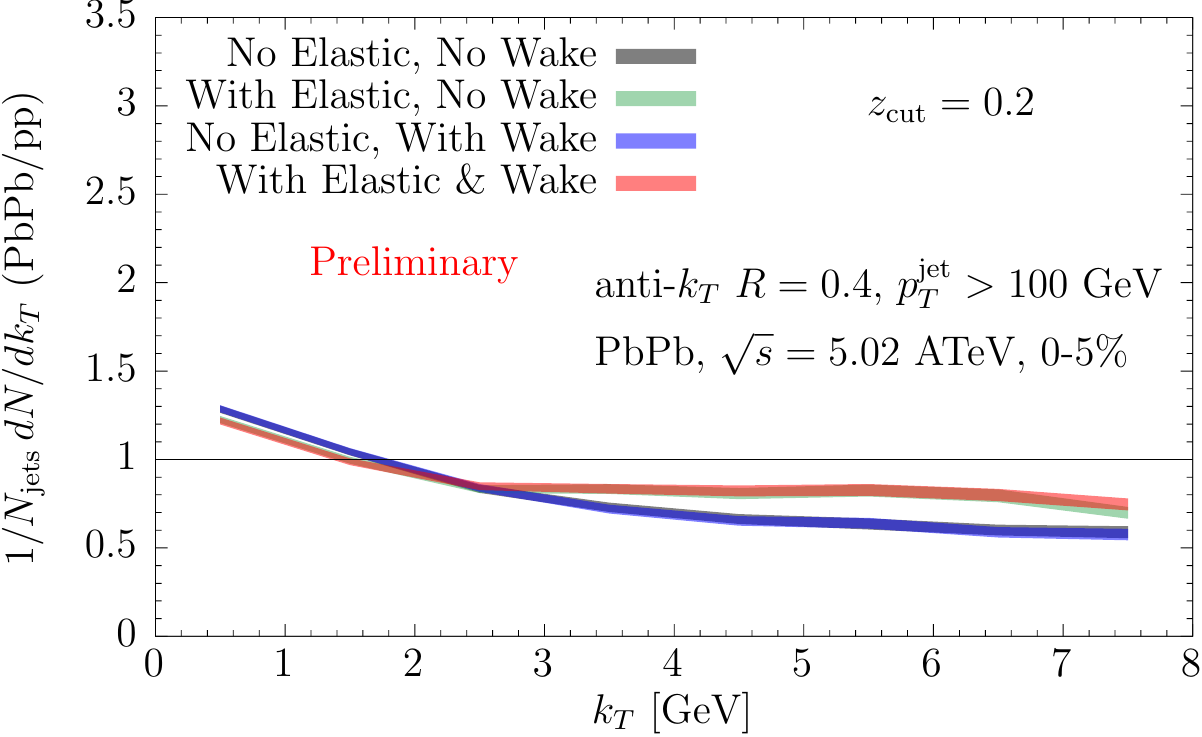}
\caption{Top: 
results for two ungroomed  observables;  
fragmentation function (top-left) and jet shape (top-right).
Bottom: results for two groomed observables;
SoftDrop branching angle $R_g$ (bottom-left) and highest $k_T$ in the clustering tree (bottom-right).
We plot the ratio of each in PbPb and pp collisions, with/without elastic scattering,
and with/without hadrons from the wake of the jet, to see the impact of Moli\`ere scatterings separately and in combination with that of the wake.}
\label{Fig:groomed-and-ungroomed}
\end{figure}

We begin by looking at observables for jets reconstructed with the anti-$k_t$ algorithm with $R=0.4$.  
Since Moli\`ere scattering kicks some energy outside $R=0.4$,
it increases the suppression $R_{AA}$ of jets and high-$p_T$ hadrons in PbPb collisions,
and we find that the fitted $\kappa_{\rm sc}$ is reduced by $\sim 10$\%.
In Figs.~2 and 3 
we show results obtained without wake and without elastic scatterings in black; with elastic scatterings and no wake in green; without elastic scatterings but with wake in blue; and with both effects included, as in experimental measurements, in red. In Fig.~2 (upper panels)  we show two standard ungroomed jet observables: the jet fragmentation function which describes the distribution in longitudinal momentum fraction $z$ of particles within the jet and the jet shape
which describes the distribution of jet energy as a function of angle $r$ relative to the jet axis. We see a big contribution to both observables coming from the wake
deposited in the medium by the jet -- compare red/blue to green/black.  The further effect coming from Moli\`ere scatterings is smaller in magnitude, and similar in shape (contributing most at small $z$ and larger $r$).  Elastic scattering yields only a minor enhancement to the larger effect coming from the wake, not a distinctive signature.

The groomed observables in the lower panels of Fig.~2 are more promising. 
We look at the angle $R_g$ between the jet axis and the first branch to satisfy the SoftDrop condition~\cite{Larkoski:2014wba}, and the highest $k_T\equiv z(1-z)p_T \sin\theta$ in the clustering tree.
For both of these groomed observables, red/green are well separated from blue/black, meaning that the effects of the wake on these observables are smaller in magnitude than the effects of Moli\`ere scattering.
Once we have groomed away the contribution coming from the wake, we see the enhancement of semi-hard structures at moderate-to-large angles coming from Moli\`ere scatterings kicking partons within a jet.

\begin{figure}[t!]
\centering
\includegraphics[width=0.49\textwidth]{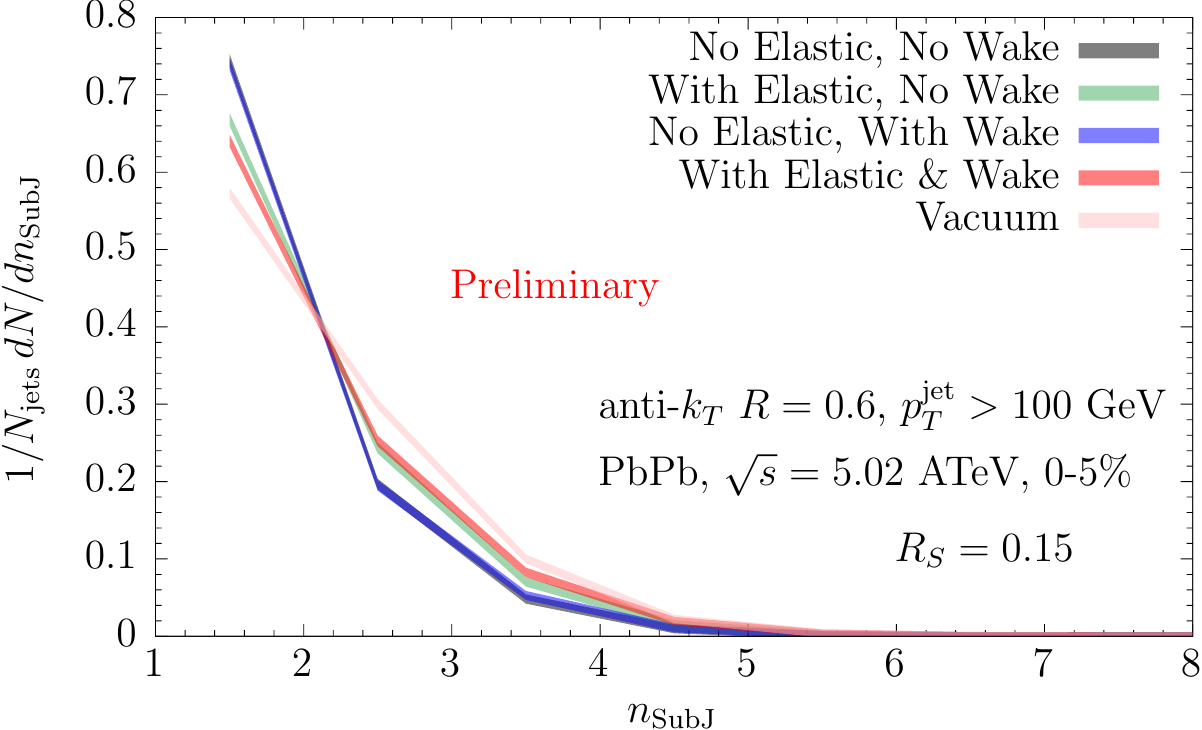}
\includegraphics[width=0.49\textwidth]{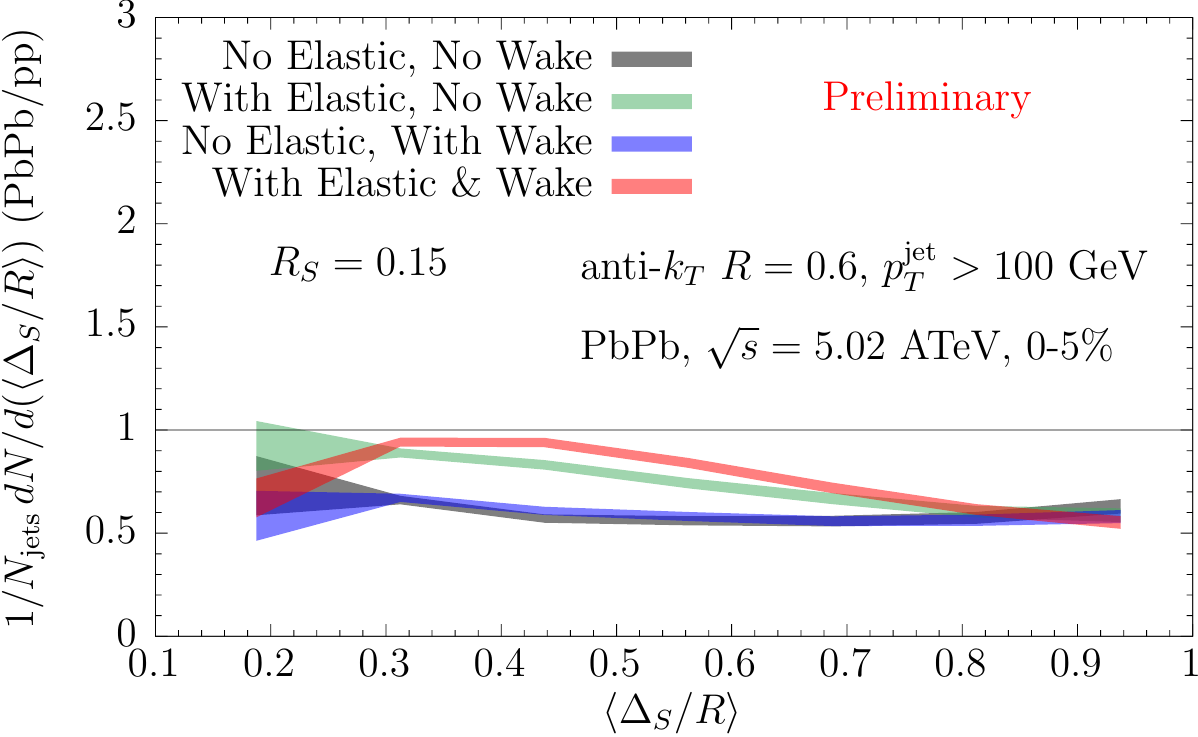}
\caption{Results for the properties of the inclusive subjets distributions analyzing the impact of the elastic scatterings separately and in combination, for the subjet multiplicity (left panel) and the average angular distance among them (right panel).}
\label{Fig:substruct}
\end{figure}

The observables in Fig.~3 are promising, but once we understand what we are looking for we can design observables for the purpose of seeing the effects of Moli\`ere scattering of partons within a jet shower.
An example is shown in Fig.~\ref{Fig:substruct}, where we look at ``subjets within jets''. We first reconstruct jets with $R=0.6$ and then re-run the anti-$k_T$ algorithim with $R=0.1$ on the particles in an $R=0.6$ jet.  This allows us to look for a jet ``sprouting additional prongs'' due to Moli\`ere scattering of shower partons.
We see in Fig.~3 that turning on Moli\`ere scattering increases the multiplicity of subjets (left panel) and the average angular distance among them $\langle \Delta_S \rangle$ (right panel). 
Mol\`ere scattering increases number of subjets within jets (although they are still fewer than in vacuum due to the omnipresent selection bias effect towards narrower jets~\cite{Casalderrey-Solana:2019ubu})
and widens their angular distribution.

The results in these proceedings are preliminary, as explained in the previous Section; their completion is ongoing and will be presented in an upcoming publication.
We have looked at other observables, 
for example finding to date that $Z$-jet and hadron-jet acoplanarity observables are more affected by the wake than by Moli\`ere scattering,
and shall report upon these investigations in our publication to come. 
It would be interesting in future work to investigate acoplanarity observables based upon other ways of defining the jet direction, including for example by focusing on charmed hadrons within jets.
Subjets-within-jets observables as in Fig.~3 are the best observables that we have found to date
for seeing the effects of Moli\`ere scattering of jet partons off quasiparticles in QGP.

\section*{Acknowledgements}
DP has received funding from the European Union's Horizon 2020 research and innovation program under the Marie Skłodowska-Curie grant agreement No. 754496. The work of KR was supported by the U.S. Department of Energy, Office of Science, Office of Nuclear Physics grant DE-SC0011090.

\bibliographystyle{ieeetr}
\bibliography{mybibfile}

\end{document}